\documentclass[11pt]{article}
\usepackage[utf8]{inputenc}
\usepackage[T1]{fontenc}
\usepackage[a4paper,margin=2.5cm]{geometry}
\usepackage{amsmath,amssymb}
\usepackage{mathrsfs}
\usepackage{booktabs}
\usepackage{tabularx}
\usepackage{float}
\usepackage[hidelinks]{hyperref}

\newcolumntype{L}{>{\raggedright\arraybackslash}X}
\DeclareMathOperator{\grad}{grad}
\renewcommand{\div}{\operatorname{div}}
\DeclareMathOperator{\tr}{tr}

\title{Stroh--Hamiltonian Framework for Modeling Incompressible Poroelastic Materials\thanks{Published in \emph{Mathematics} \textbf{2025}, \emph{13}(9), 1436. \url{https://doi.org/10.3390/math13091436}}}
\author{Kinza Arshad\thanks{These authors contributed equally to this work. E-mail: karshad@unisa.it} \and Vincenzo Tibullo\thanks{These authors contributed equally to this work. Corresponding author. E-mail: vtibullo@unisa.it}}
\date{Department of Mathematics, University of Salerno, 84084 Fisciano, Italy}

\begin{document}

\maketitle

\begin{abstract}
The Stroh sextic formalism, developed by Stroh, offers a compelling framework for representing the equilibrium equations in anisotropic elasticity. This approach has proven particularly effective for studying multilayered structures and time-harmonic problems, owing to its ability to seamlessly integrate physical constraints into the analysis. By recognizing that the Stroh formalism aligns with the canonical Hamiltonian structure, this work extends its application to Biot's poroelasticity, focusing on scenarios where the solid material is incompressible and there is no fluid pressure gradient. The study introduces a novel Hamiltonian-based approach to analyze such systems, offering deeper insights into the interplay between solid incompressibility and fluid–solid coupling. A key novelty lies in the derivation of canonical equations under these constraints, enabling clearer interpretations of reversible poroelastic behavior. However, the framework assumes perfectly drained conditions and neglects dissipative effects, which limits its applicability to fully realistic scenarios involving energy loss or complex fluid dynamics. Despite this limitation, the work provides a foundational step toward understanding constrained poroelastic systems and paves the way for future extensions to more general cases, including dissipative and nonlinear regimes.
\end{abstract}

\noindent\textbf{Keywords:} Stroh formalism; Biot's poroelasticity; Hamiltonian formalism; solid incompressibility

\noindent\textbf{MSC 2020:} 74A05; 74A10; 74A60

\section{Introduction}\label{sec1}

The Stroh formalism, introduced by A.N. Stroh~\cite{stroh1962steady}, is a key concept in the study of anisotropic elasticity. It establishes a coherent framework for analyzing wave propagation and stress fields in materials with complex, directionally dependent properties~\cite{hwu2022stroh}. It is also a foundational method for studying elastic materials with general constraints and has been used to examine the effect of pre-stress on surface waves in an inextensible material~\cite{edmondson2009stroh, chadwick1997application}. Its utility stems from its ability to not only handle the complexities of anisotropic elasticity but also integrate physical constraints seamlessly into the analysis, making it a foundational tool for researchers and engineers working in material science, mechanics, and wave propagation studies~\cite{wang2024study, qu2024defects, wu2000extension, ensam2023transmission, pan2023lamb}. 

Poroelastic materials are extensively utilized across various fields of physics due to their exceptional dissipative properties~\cite{farm2016absorption, wang2024characteristics, yang2017sound, benjamin2024impacts}. Particularly,  Biot’s theory of poroelasticity has established itself as a remarkably effective phenomenological model, with wide-ranging applications across numerous scientific and engineering fields. Specifically, in automotive applications, trim components absorb air, altering the system's dynamic behavior, which is crucial for Noise, Vibration, and Harshness (NVH) studies. The Biot poroelastic model enables the precise frequency response analysis of trim materials, carpets, and foams, improving solution accuracy. One prominent area of investigation has been the behavior of wave propagation in porous media, a subject comprehensively examined in the review by~\cite{corapcioglu1996propagation}. The theory itself has undergone significant extensions to address more complex phenomena and broaden its applicability. For instance, researchers have incorporated concepts such as double-porosity and multi-porosity systems, which are crucial for modeling materials with hierarchical pore structures~\cite{berryman2000elastic, pramanik2024theory}. Further, the theory has been expanded to include piezoelectric effects, as explored by~\cite{sharma2010piezoelectric}, which opens new avenues for applications in smart materials and sensors.  Recently, ref.~\cite{chen2018biot} proposed a Biot's model for deriving material properties of cancellous bone from acoustic measurements. He modified Biot's equations for cancellous bone and coupled them with a boundary integral equation for water pressure.
Biot's poroelasticity also has widespread application when it comes to poroelastic rocks~\cite{shao1998poroelastic,yang2024modified}. Rocks are naturally filled with cracks and pores that are often saturated with one or more fluid phases. These features are crucial in many challenges faced in rock mechanics, petroleum engineering, geophysics, and other fields, where the presence of cracks and discontinuities in rock formations affects material behavior. As such, it is essential to consider the effects of the porous medium in these problems, as they significantly influence fluid flow, stress distribution, and overall rock performance. In this context, ref.~\cite{abdollahipour2016complete} studied Biot's original formulation of poroelasticity, which included the solid and fluid components of geomaterials, and presented a complete formulation of an indirect boundary element method for poroelastic rocks. 
{Despite these advancements, one notable gap in the literature is the absence of a Stroh-like formalism for poroelasticity. Such a framework, widely used in other areas of mechanics, has yet to be adapted or developed for multi-field theories like poroelasticity. This absence may be attributed to the intrinsic complexity of multi-field systems, which pose significant challenges to the adoption of a Stroh-like approach.
Stroh's method is often considered a ``Hamiltonian'' approach to elasticity, as it utilizes concepts similar to those in Hamiltonian mechanics. This Hamiltonian nature of the Stroh formulation has been observed in passing by several researchers in earlier works, for example, see~\cite{barnett2000bulk, fu2007hamiltonian, nobili2022hamiltonian}. This approach addresses a significant gap in the literature and offers a structured method to study constrained poroelastic systems and highlight the distinctive behavior of the solid phase under incompressibility conditions, where fluid–solid interactions significantly influence stress distribution and deformation.
Nobili's earlier work~\cite{nobili2024hamiltonian} primarily developed a Stroh-like formulation for reversible poroelasticity under perfectly drained conditions, which also applied in a similar manner to thermoelasticity in perfect conductors~\cite{nobili2021quasi}. Along with this, he addressed the compressible behavior of both the fluid and solid phases, with a brief acknowledgment of incompressibility as a limiting scenario. However, a detailed derivation for the incompressibility condition, particularly for the solid phase, has not been explored. In this study, we started with the incompressibility constraint for the solid phase using the observations of Fu~\cite{fu2007hamiltonian} and derived the canonical equations using the Hamiltonian density framework as described by Nobili~\cite{nobili2024hamiltonian}. It can help accurately model the mechanical response of the material in multilayered solids, composite materials, and time-harmonic problems. This is important for many applications in fields that span diverse areas, including geophysics (e.g., seismic wave propagation), biomechanics (e.g., tissue deformation), material science (e.g., smart or hierarchical materials), and engineering (e.g., testing mechanical properties of constrained systems).

The rest of the article has been organized in the following manner. Section~\ref{sec2} presents a review of reversible poroelasticty. Section~\ref{sec3} deals with the Hamiltonian formalism and derivation of the canonical equation for the present problem. A few concluding remarks are included in Section~\ref{sec4}.

\section{Reversible Poroelasticity Under Incompressibility of Solid in a Poroelastic Material}\label{sec2}

Consider a two-dimensional linear, steady poroelastic material with the seepage displacement, describing the fluid's motion relative to the solid per unit volume of the poroelastic medium, defined as follows (see, e.g., Table~\ref{tab:notation} for all key {variables):} \begin{equation}
    \mathbf{w}=f(\mathbf{U}-\mathbf{u})
\end{equation}
{Here,} \(\mathbf{u}\) and \(\mathbf{U}\) are considered to be the displacements of the solid and fluid phases, respectively.  \(f\) represents the effective porosity, which is non-uniform and characterizes the interconnected pore space.
We further assume that  
\begin{align*}
    e=\div\mathbf{u}, \quad \zeta=-\div\mathbf{w}
\end{align*}
where \(e\) represents the volume increment in the solid and \(\zeta\) volume increment in the fluid phase. Because of an incompressible solid material under consideration, we have an additional condition~\cite{fu2007hamiltonian}
\begin{align}
   u_{i,i}= \div\mathbf{u}=e=0 
\end{align}
Furthermore, we have an important relation~\cite{biot1962mechanics}
\begin{align}
    -\zeta=(\mathbf{U-u}) \cdot \grad  f+f\epsilon 
\end{align}
where \( \epsilon =\div\mathbf{U}\). In case of uniform porosity along with the condition of incompressible solid material, we have the following result~\cite{biot1962mechanics}: 
\begin{align*}
\epsilon=-f^{-1}\zeta
\end{align*}
\begin{table}[htbp]
\centering

\caption{List of notations and symbols used in the Stroh–Hamiltonian framework.}
\label{tab:notation}

\begin{tabularx}{\textwidth}{lL}
\toprule
\textbf{Symbol} & \textbf{Description} \\
\midrule
$ \mathbf{u} $ & Displacement field of the solid phase  \\
$ \mathbf{U} $ & Displacement field of the fluid phase  \\
$ \mathbf{w} $ & Seepage displacement ($ \mathbf{w} = f(\mathbf{U} - \mathbf{u}) $) \\
$ f $ & Effective porosity (non-uniform)  \\
$ e $ & Volumetric strain of the solid ($ e = \div \mathbf{u} $)  \\
$ \zeta $ & Volumetric strain of the fluid ($ \zeta = -\div \mathbf{w} $)  \\
$ \epsilon $ & Volumetric strain of the fluid phase ($ \epsilon =  \div   U $)  \\
$ \mathbf{T} $ & Total stress tensor  \\
$ \boldsymbol{\sigma} $ & Effective stress tensor in the solid phase  \\
$ p_f $ & Fluid pressure (positive in compression)  \\
$ \alpha $ & Effective stress coefficient ($ \alpha = r/m $)  \\
$ m $ & Compressibility modulus of the fluid  \\
$ \mathbf{E} $ &  Linear deformation tensor \\
$ r $ & Cross-coupling term between solid and fluid phases \\
$ \tilde{\mathbb{C}} $ & Elastic moduli tensor of the solid skeleton  \\
$ Q, R, T $ & Stroh matrices  \\
$ \mathbf{t}_1, \mathbf{t}_2 $ & Fundamental force vectors \\ %

$ L $ & Lagrangian density \\
$ H $ & Hamiltonian density  \\ 
$ \lambda $ & Positive constant ensuring positive definiteness of $ \hat{T} $  \\ 
$ \hat{Q}, \hat{R}, \hat{T} $ & Modified Stroh matrices  \\
$ \mathcal{L} $ & Total energy functional \\ 
$ W $ & Stored potential energy density  \\ 
$ p $ & Lagrange multiplier enforcing incompressibility \\
\bottomrule
\end{tabularx}

\end{table}

Let us assume \((O, x_1,x_2,x_3)\) is an orthogonal reference frame \( \mathscr{E}\) and \(\mathbf{n}\) is the unit vector normal to any relevant directed surface \(S\). Moreover, an orthonormal set of basis vectors \(\mathbf{e}_1, \mathbf{e}_2\), and \(\mathbf{e}_3\)  is introduced along with axis \( (x_1,x_2,x_3) \) such that \(\mathbf{e}_i \cdot \mathbf{e}_j = \delta_{ij}\), (\(\delta_{ij}\) denotes the Kronecker delta), assuming the conventional understanding that repeated subscripts are summed over the set \(1,2,3\) and that a dot between two vectors denotes their inner product. There is also a distinction in notation between a tensor, like the identity tensor \(\mathbf{1}\), and its matrix representation, denoted \(I\), with components given by \(\mathbf{1}_{ij} = \delta_{ij}\) \cite{nobili2024hamiltonian}.

Total stress denoted by the rank-2 symmetric tensor \(\mathbf{T}\) is defined as
\begin{align}
    \mathbf{T}=\boldsymbol{\sigma} -\alpha p_f \mathbf{1}- p \mathbf{1}    \label{4} \end{align}

Here, the combination of the effective stress in the solid phase, \(\boldsymbol{\sigma}\), and the stress in the fluid phase, \(-p_f \mathbf{1}\), is made through the effective stress coefficient \(\alpha\) along with the Lagrange multiplier \( p\) in order to enforce the constraint, owing to the fact that the solid is incompressible \( (\div\mathbf{u} = 0 ) \) \cite{berger1999stabilized, abousleiman2002porothermoelasticity}.
Here, \(p_f\) is the fluid pressure (positive for compressive stresses) per unit area of the fluid phase. The following applies to an anisotropic solid skeleton:
\begin{align}
    \boldsymbol{\sigma}= \tilde{\mathbb{C}} \mathbf{E} \label{5}
\end{align}
where \(\tilde{\mathbb{C}}\) is the rank-4 tensors of elastic moduli whose components in the reference frame 
\( \mathscr{E}\) are denoted by  \(\tilde{c}_{ijkl}\),
 and \(\mathbf {E}= \frac{1}{2}(\grad\mathbf{u}+ \grad\mathbf{u}^T)\) is the linear deformation tensor,  such that \(e=\tr \mathbf{E}=\div \mathbf{u}\). For a poroelastic material with an incompressible solid~\cite{nobili2024hamiltonian}
 
\begin{align}
     p_f = m \zeta  \label{6}  \end{align}
 where \(m\) is the compressibility modulus of fluid, i.e., 
 the fluid pressure required to force a unit volume of fluid into the pore structure while keeping the solid volume unchanged%
 . Also, following~\cite{schanz2009poroelastodynamics}
\begin{align}
     f\leq \alpha  =r/m=1
 \end{align}
 where \(r\) represents the cross coupling term between volume changes in the solid and in the fluid. 
 Putting Equations \eqref{5} and \eqref{6} into the total stress \eqref{4} and simplifying,  we obtain
\begin{align}
 \mathbf{T} =\tilde{\mathbb{C}}\mathbf{E}-r \zeta \mathbf{1} - p \mathbf{1} \end{align}
We now define a fundamental force vector owing to the fact that we have an incompressible anisotropic solid, following~\cite{nobili2024hamiltonian, fu2007hamiltonian}
\begin{equation}
    \label{9}
    \begin{aligned}
        & \mathbf{t}_1 = \mathbf{T}\mathbf{e}_1 = Q  \mathbf{u}_{,1} + R \mathbf{u}_{,2} - r\zeta \mathbf{e}_1 - p\mathbf{e}_1   \\
        & \mathbf{t}_2 = \mathbf{T}\mathbf{e}_2 = R^T\mathbf{u}_{,1} + T \mathbf{u}_{,2} - r\zeta \mathbf{e}_2 - p\mathbf{e}_2 
    \end{aligned}   
    \end{equation}
where \(Q_{ij}=c_{i1j1}\), \(R_{ij}=c_{i1j2}\), and \(T_{ij}=c_{i2j2}\) are the usual Stroh matrices and a subscript comma denotes differentiation, e.g.,~\(\mathbf{u}_{,1}=\partial\mathbf{u}/\partial x_1\). In particular, \(Q\) and \(T\) are symmetric and positive definite, provided the strain energy is a positive function. \( p \) is a Lagrange multiplier introduced to permit independent variation of \( \mathbf{u}_{,1} \) and \( \mathbf{u}_{,2} \).

It is crucial to note that the reversibility assumptions as observed by~\cite{biot1957elastic}, encapsulated by the stored elastic potential \( W \), allow for the identification of the coupling coefficient in the third terms of Equation \eqref{9} with the corresponding term in \eqref{6}. Strong ellipticity stipulates that \( Q \) and \( T \) must be positive definite, and~\( m > 0 \). For convenience, we adopt an isotropic approximation for the cross-coupling term; for a comprehensive treatment of transverse anisotropy, see \cite{biot1955theory}. Moreover, we assume that the dependent variables are invariant with respect to \( x_3 \), i.e.,~\( \partial / \partial x_3() = 0 \).

Next, we write the potential elastic energy minus the work carried out by the applied external forces over the body \( B \)  
(which means the total energy in the sense of Eshelby)
\begin{align}
 \mathcal{L} = \int_B W dV-\int_{\partial B} (\mathbf{t}_0\cdot \mathbf{u}-p_{f_0} \mathbf{n}\cdot \mathbf{w}) dS   \label{10}
\end{align}
where we assumed the stored potential energy density to be of the form
\begin{align}
    W = \frac{1}{2} (\mathbf{T} \cdot \grad\, \mathbf{u} + p_f \zeta)
\end{align} 
Here, \(\mathbf{t}_0\) and \(p_{f_0}\) are the prescribed surface force and fluid pressure over the body boundary \(\partial B\) with the unit normal \(\mathbf{n}\). For the sake of simplicity, no body force is considered. 
Taking into account \eqref{4}, we are able to write
\begin{align}
   W = \frac{1}{2}(\boldsymbol{\sigma} \cdot \mathbf{E} + p_f \zeta)   
\end{align}
or
\begin{align}
     W = \frac{1}{2}(\boldsymbol{\sigma} \cdot \mathbf{E} + m \zeta^2)
\end{align}

In the context of a reversible process (i.e., thermostatics), the imposition of boundary pressure should not induce any movement within the fluid phase. As a result, \( p_{f_0} \) remains constant on the boundary surface \( \partial B \), and~\( p_f \) is uniformly constant throughout the material~\cite{biot1962mechanics}. This is referred to as the perfectly drained condition, which ensures that the fluid pressure does not change during mechanical deformation. This uniform pressure distribution holds during the steady-state motion of the solid, assuming dissipation is not considered. The negative sign of the fluid pressure reflects its positive value under compression. In~\eqref{10}, we assume 
\begin{align*}
    \mathbf{T} \cdot \grad \,{\mathbf{u}}= \mathbf{t}_1 \cdot \mathbf{u}_{,1}+ \mathbf{t}_2 \cdot \mathbf{u}_{,2}
\end{align*}
where it is already clear that 
\begin{align*}
    \grad \, \mathbf{u} = \mathbf{u}_{,1} \otimes  \mathbf{e}_1 +\mathbf{u}_{,2} \otimes  \mathbf{e}_2.
\end{align*} 

The equilibrium equation {reads} 
\begin{subequations}
\begin{align}
&    \mathbf{t}_{1,1}+ \mathbf{t}_{2,2}=\mathbf{0} \label{14 a} \\
 &    \grad p_f = \mathbf{0} \implies \grad(m \zeta) = \mathbf{0} \label{14 b} \end{align} 
\end{subequations}
where the last equation  is the equilibrium equation of Darcy's law~\cite{biot1962mechanics}. Without a loss of generality, we assume that the boundary conditions in terms of forces can be expressed as 
\begin{subequations}
\begin{align}
&    \mathbf{T n }= \mathbf{t}_0,  \quad p_f=p_{f0} \quad  \text{for} \quad x \in \partial B  \label{15 a} \\
& \mathbf{u}=\mathbf{0}\quad  \text{on} \quad \partial B   \label{15 b}  
\end{align}    
\end{subequations}
where \(\partial B \) is the boundary of the body \(B\). By the divergence theorem, we can rewrite the total energy
\begin{align}
    \mathcal{L}=-\int_B L dV 
\end{align}
where \( L \) is the Lagrange density given by
\begin{align}
  L(\mathbf{u}_{,1}, \mathbf{u}_{,2}, \mathbf{w}_{,1}, \mathbf{w}_{,2}) =\frac{1}{2} \mathbf{T} \cdot \grad\mathbf{u} +\frac{1}{2}p_f \zeta - (r \zeta + \frac{1}{2}p)\mathbf{u}_{i,i}
\end{align}
the last term counterbalances the effect of the solid's incompressibility, given that the fluid phase is compressible in poroelastic material. 
Using \eqref{6} along with the definition of fundamental force vectors in (9), we obtain
\begin{align}
L = \frac{1}{2} \mathbf{u}_{,1} \cdot Q \mathbf{u}_{,1}+\mathbf{u}_{,1} \cdot R \mathbf{u}_{,2}+\frac{1}{2}\mathbf{u}_{,2} \cdot  T \mathbf{u}_{,2}+\frac{1}{2}m \zeta^2-p\mathbf{u}_{i,i} - r\zeta\mathbf{u}_{i,i} \end{align} 
Since the formulation admits the Stroh formalism, the Lagrangian density becomes
\begin{align}
L=    \frac{1}{2} (\mathbf{u}_{,1} \cdot Q \mathbf{u}_{,1}+\mathbf{u}_{,1} \cdot R \mathbf{u}_{,2}+\frac{1}{2}\mathbf{u}_{,2} \cdot  T \mathbf{u}_{,2})+\frac{1}{2}m (\mathbf{e}_1 \cdot \mathbf{w}_{,1} +\mathbf{e}_2 \cdot \mathbf{w}_{,2}) ^2-p\mathbf{u}_{i,i} - r \zeta\mathbf{u}_{i,i}
\end{align}
As the Euler–Lagrange equation is defined as
\begin{subequations}
\begin{align}
 &  \frac{d}{dx_1}\frac{\partial L}{ \partial \mathbf{u}_{,1}}+\frac{d}{dx_2}\frac{\partial L}{ \partial \mathbf{u}_{,2}}=0,  \label{eq:21a} \\
 & \frac{d}{dx_1}\frac{\partial L}{ \partial \mathbf{w}_{,1}}+\frac{d}{dx_2}\frac{\partial L}{ \partial \mathbf{w}_{,2}}=0 \label{21 b}
 \end{align}
\end{subequations}
This takes the form of a Stroh formalism once we identify either coordinate, say \(x_2\), as the time-like variable, e.g, see~\cite{fu2007hamiltonian}; thus, \eqref{eq:21a} gives
\begin{align}
   \frac{1}{2}(2Q \mathbf{u}_{,1}+R \mathbf{u}_{,2}-p\mathbf{e}_1- r \zeta\mathbf{e}_1 )_{,1}+(\mathbf{u}_{,1}\cdot R +T \mathbf{u}_{,2}-p \mathbf{e}_2 -r \zeta \mathbf{e}_2 )_{,2}=0
 \end{align}
 that corresponds to the equilibrium Equation \eqref{15 a}, provided that we account for (9). Similarly, \eqref{21 b} becomes
\begin{align}
(m \zeta)_{,1}\mathbf{e}_1+(m \zeta) _{,2} \mathbf{e}_2=0  
\end{align} 
This aligns with \eqref{15 b}, once  \eqref{6} is acknowledged.

\section{Hamiltonian Formalism}\label{sec3}
To introduce the Hamiltonian formalism, we treat \(x_2\) as a time-like variable~\cite{fu2007hamiltonian}. As a result, a superscript dot will be used to denote differentiation with respect to \(x_2\). We take into account
\begin{align}
    \hat{Q}= Q+\lambda \mathbf{e}_1 \otimes \mathbf{e}_2,  \quad \hat{R}=R+\lambda \mathbf{e}_1 \otimes \mathbf{e}_2 ,\quad \hat{T}=T+ \lambda \mathbf{e}_2 \otimes \mathbf{e}_2 
    \end{align}
where \(\lambda\) is a positive constant that is large enough introduced here because for incompressible materials, the stiffness matrix associated with the volumetric deformation becomes singular because the material cannot sustain hydrostatic stresses. To avoid this singularity, \( \lambda\) is added to the diagonal terms of the stiffness matrices. This ensures that \(\hat{T}\) remains positive definite, which is essential for the stability of numerical computations and the invertibility of the matrices~\cite{fu2007hamiltonian}, 
 whence we may rewrite \eqref{9} as 
\begin{subequations}
\begin{align}
&        \mathbf{t}_1= \hat{Q}\mathbf{u}_{,1}+\hat{R}\dot{\mathbf{u}}-r \frac{p_f}{m} \mathbf{e}_1 -p \mathbf{e}_1\\
&        \mathbf{t}_2=\hat{R}^T \mathbf{u}_{,1}+\hat{T}\dot{\mathbf{u}}-r\frac{p_f}{m}\mathbf{e}_2 -p \mathbf{e}_2  
\end{align}
\end{subequations}
and thus the Lagrangian density becomes
\begin{align}
 L  = \frac{1}{2} \mathbf{u}_{,1} \hat{Q} \mathbf{u}_{,1}+ \mathbf{u}_{,1} \hat{R}\dot{\mathbf{u}} +\frac{1}{2}\dot{\mathbf{u}} \cdot \hat{T} \dot{\mathbf{u}} + \frac{1}{2}m(\mathbf{e}_1\cdot  \mathbf{w}_{,1}+\mathbf{e}_2 \cdot \mathbf{w}_{,2})^2-p\mathbf{u}_{i,i} - r \zeta\mathbf{u}_{i,i}
\end{align} 
and hence the conjugate momenta are obtained as
\begin{align}
 &   \mathbf{p}_1= \frac{\partial L}{\partial \dot{\mathbf{u}}}= \mathbf{u}_{,1}\hat{R}+\hat{T} \dot{\mathbf{u}}-p \mathbf{e}_2  - r \zeta \mathbf{e}_2=\mathbf{t}_2  \\
 & \mathbf{p}_2 = \frac{\partial L}{\partial \dot{\mathbf{w}}}= -  p_f \mathbf{e}_2 
\end{align}
where 
\begin{align*}
    p_f = \zeta m 
\end{align*}
solving (26) for \(\dot{\mathbf{u}}\) leads to
\begin{align}
     \dot{\mathbf{u}}= \hat{T}^{-1}(\mathbf{t}_2-\hat{R}^T \mathbf{u}_{,1} + r \frac{p_f}{m} \mathbf{e}_2 +p \mathbf{e}_2) \label{28}
 \end{align}
Via the scalar multiplication of \eqref{28} by \(\mathbf{e}_2\)
\begin{align*}
 &   \dot{\mathbf{u}}\cdot \mathbf{e}_2= \hat{T}^{-1}(\mathbf{t}_2 -\hat{R}^T \mathbf{u}_{,1}+\frac{r}{m}p_f \mathbf{e}_2 +p \mathbf{e}_2) \cdot \mathbf{e}_2 \\ \end{align*}
By using \( \mathbf{u}_{,1} \cdot \mathbf{e}_1 +\mathbf{u}_{,2} \cdot \mathbf{e}_2 =0\) and eliminating \(p\) following~\cite{fu2007hamiltonian}:
\begin{align}
    p= \zeta_1 (\hat{T}^{-1}(\mathbf{t}_2 -\hat{R}^T \mathbf{u}_{,1})+ r \zeta -\mathbf{u}_{,1}\cdot \mathbf{e}_1) \end{align} 
where
\begin{align}
    \zeta_1= \frac{1}{\mathbf{e}_2 \cdot \hat{T}^{-1} \mathbf{e}_2 } \end{align} 
Reflecting on \(\zeta=-\div\mathbf{w}\), we have
\begin{align}
   - \dot{\mathbf{w}} \cdot \mathbf{e}_2  = \mathbf{w_{,1}} \cdot \mathbf{e}_1 +\frac{p_f}{m} \label{31}
\end{align}
In this framework, we define the Hamiltonian density.
\begin{align}
&    H = \mathbf{t}_2 \cdot \dot{\mathbf{u}} + \mathbf{p}_2 \cdot \dot{\mathbf{w}}-L  \\
 & \nonumber    = \mathbf{t}_2 \cdot \dot{\mathbf{u}} - \mathbf{p}_2 \cdot \left(\mathbf{w}_{,1} \cdot \mathbf{e}_1 +\frac{p_f}{m}\right) - \Big(\frac{1}{2}\mathbf{u}_{,1}\hat{Q}\mathbf{u}_{,1}+\mathbf{u}_{,1}\hat{R}\dot{\mathbf{u}} +\frac{1}{2}\dot{\mathbf{u}} \cdot \hat{T} \dot{\mathbf{u}} + \\
 &  + \frac{1}{2}m(\mathbf{e}_1\cdot  \mathbf{w}_{,1}+\mathbf{e}_2 \cdot \mathbf{w}_{,2})^2-p\mathbf{u}_{i,i} - r \zeta\mathbf{u}_{i,i}\Big) \end{align}
 by further simplifying H, we obtain
\begin{align}
& \nonumber H  = \frac{1}{2}( \mathbf{t}_2 -\hat{R} \mathbf{u}_{,1} +p \mathbf{e}_2 +r \zeta \mathbf{e}_2) \hat{T}^{-1}( \mathbf{t}_2 -\hat{R} \mathbf{u}_{,1} +p \mathbf{e}_2 +r \zeta \mathbf{e}_2)  -\frac{1}{2}\mathbf{u}_{,1}Q \mathbf{u}_{,1} - \\
& - \frac{1}{2}m \zeta^2 p \mathbf{u}_{,1}\mathbf{e}_1+r \zeta \mathbf{u}_{,1}\mathbf{e}_1 - p_f \left( \mathbf{w}_{,1} \cdot \mathbf{e}_1 + \frac{p_f}{m}\right)
\end{align}
Canonical equations, by definition, can be classified into two categories, which are commonly expressed as a vector equation
\begin{align}
    \mathbf{\dot{q}}= \frac{\partial H}{\partial \mathbf{p}} \quad \text{and} \quad \mathbf{\dot{p}}= -\frac{\partial H}{\partial \mathbf{q}}
\end{align}
In the first group, we have
\begin{align} \label{eq 36}
 \dot{\mathbf{u}}= \frac{\partial H}{ \partial \mathbf{t}_2}= \hat{T}^{-1}(\mathbf{t}_2 -\hat{R}\mathbf{u}_{,1}+p \mathbf{e}_2 +r \zeta \mathbf{e}_2) =\mathbf{t}_2 
\end{align}
 and
\begin{align} \label{eq 37}
&  \nonumber   \dot{\mathbf{w}} \cdot \mathbf{e}_2= -\frac{\partial H}{\partial p_f}= -\left( \frac{2}{2} \frac{r}{m} \mathbf{e}_2 \cdot \hat{T}^{-1}( \mathbf{t}_2 -\hat{R}^T \mathbf{u}_{,1}+\frac{r p_f}{m}\mathbf{e}_2+p\mathbf{e}_2)-\frac{p_f}{m}+\frac{r}{m}\mathbf{u}_{,1}\cdot \mathbf{e}_1 \right)\\
& = -\left(\mathbf{w}_{,1}\cdot \mathbf{e}_1 + \frac{p_f}{m} \right) 
\end{align}
that correspond to \eqref{28} and \eqref{31}, respectively. 

The second group yields the equilibrium equation. Specifically, it takes the form:
\begin{align}\label{eq 38}
 \dot{\mathbf{t}}_2= \frac{\partial H}{ \partial \mathbf{u}}= -\left(\hat{R} \hat{T}^{-1}(\mathbf{t}_2-\hat{R}^T \mathbf{u}_{,1}+\frac{r p_f}{m}\mathbf{e}_2 +p \mathbf{e}_2 ) +\hat{Q}\mathbf{u}_{,1}- p \mathbf{e}_1
- r \frac{p_f}{m}\mathbf{e}_1\right)_{,1}
\end{align} 
that accounts for \eqref{28}%
, whereby \(\hat{T}^{-1}\) times the term in round brackets gives \(\dot{\mathbf{u}}\), and in light of first of \eqref{9}, amounts to \eqref{14 a}. By the same token,
\begin{align}\label{eq 39}
    -\dot{p}_f\mathbf{e}_2= -\frac{\partial H}{ \partial \mathbf{w}}=(p_f \mathbf{e}_1)_{,1}
\end{align}
that is, immediately%
. 

{This} work demonstrates the effectiveness of the Stroh–Hamiltonian formalism in analyzing wave propagation, stress distributions, and dynamic responses in incompressible poroelastic materials. For instance, this  can be used to analyze wave propagation in an incompressible poroelastic material under perfectly drained conditions for the dispersion relation of compressional and shear waves using the canonical equations. The framework may reproduce the known behavior of wave speeds and attenuation in incompressible materials while highlighting the role of fluid–solid coupling. Moreover, we can also use this framework to model stress distribution in multilayered structures with incompressible solid phases subject to external loading and analyze the stress and deformation fields within each layer. This highlights the framework’s capability to handle complex geometries and material configurations, providing insights into the interplay between fluid pressure and solid deformation. 

\section{Conservation and Integral Representation of Hamiltonian Density}\label{sec4}
To determine whether the Hamiltonian density is conservative or not, we analyze its structure and dependence on the variables.
The Hamiltonian density \(H \) is given as:
\begin{align}
& \nonumber H  = \frac{1}{2}( \mathbf{t}_2 -\hat{R} \mathbf{u}_{,1} +p \mathbf{e}_2 +r \zeta \mathbf{e}_2) \cdot \hat{T}^{-1}( \mathbf{t}_2 -\hat{R} \mathbf{u}_{,1} +p \mathbf{e}_2 +r \zeta \mathbf{e}_2)  -\frac{1}{2}\mathbf{u}_{,1}\cdot Q \mathbf{u}_{,1} - \\
& - \frac{1}{2}m \zeta^2 p \mathbf{u}_{,1}\cdot\mathbf{e}_1+r \zeta \mathbf{u}_{,1}\cdot\mathbf{e}_1 - p_f \left( \mathbf{w}_{,1} \cdot \mathbf{e}_1 + \frac{p_f}{m}\right)
\end{align}
where
\begin{itemize}
    \item  \(\mathbf{t}_2\)  is a fundamental force vector,
    \item  \( \hat{Q}, \hat{R}, \hat{T} \) are modified Stroh matrices,
    \item  \(\mathbf{u}_{,1} \) and \( \mathbf{w}_{,1} \) are spatial derivatives of the displacement fields,
    \item \( \zeta = -\text{div } \mathbf{w} \) is related to fluid flow,
    \item  \( p_f = m \zeta \) is the fluid pressure,
    \item \( p \)  is a Lagrange multiplier enforcing the incompressibility constraint.
\end{itemize}

For the Hamiltonian density \( H \) to be conservative, it must satisfy the following key properties.

\subsection{Independence of \texorpdfstring{\( x_2 \)}{x2}}
From the structure of \(H \), none of the terms explicitly depend on \( x_2 \). The dependence on \(x_2 \) only enters implicitly through the variables \( \mathbf{u} \), \( \mathbf{w} \), and their derivatives. Therefore, \( H \) satisfies the first condition for energy conservation. 

\subsection{Canonical Equations}
The canonical equations derived from \( H \)  must reproduce the equilibrium equations of the system, which confirm the conservation of energy.

{The} canonical equations derived from \( H \) are evolution equations for the generalized coordinates
\[
   \dot{\mathbf{u}} = \frac{\partial H}{\partial \mathbf{t}_2}, \quad \dot{\mathbf{w}} \cdot \mathbf{e}_2 = -\frac{\partial H}{\partial p_f}.
 \]
These correspond to Equations \eqref{eq 36} and \eqref{eq 37}, which describe the evolution of the system. 

Also, for the conjugate momenta
\[
 \dot{\mathbf{t}}_2 = -\frac{\partial H}{\partial \mathbf{u}}, \quad -\dot{p}_f \mathbf{e}_2 = -\frac{\partial H} {\partial \mathbf{w}}
 \]
These correspond to Equations \eqref{eq 38} and \eqref{eq 39}, which reproduce the equilibrium equations of the system.

Since the canonical equations are satisfied, the Hamiltonian density \(H \)  correctly describes the dynamics of the system, and energy conservation is ensured.

The conservation of H reflects the reversible nature of the process, where no dissipation occurs. It has many interpretations in physics. For example, in poroelasticity the fluid can flow back and forth between pores without any loss of energy. In wave propagation, waves can travel forward or backward without attenuation. In dissipative systems, waves lose energy due to mechanisms like viscosity or friction, while in the conservative case, waves retain their amplitude and frequency indefinitely, provided no external forces act on the system. The conservation of H also provides insights into boundary effects, for instance, at free boundaries (e.g., surfaces or edges), energy can be reflected or transmitted without loss. This leads to phenomena like surface waves or edge modes, which are characteristic of conservative systems.

\subsection{Integral Representation and Symplectic Structure}
The integral representation provides a global perspective on energy conservation and reveals important properties of the system, such as its symplectic structure and connection to the Stroh formalism.  The fundamental matrix \(\hat{N} \) plays a central role in this representation~\cite{ting1996anisotropic} since \(\hat{N} \)  is symmetric:
\[
\hat{N} = \begin{bmatrix} N_3 & N_1^T \\ N_1 & N_2 \end{bmatrix},
\]
and  we let the 3-by-3 block-matrices
\begin{align}
    N_1=-\hat{T}^{-1}\hat{R}^T, \quad N_2= \hat{T}^{-1}, \quad N_3=\hat{R}\hat{T}^{-1}\hat{R}^{-1}-\hat{Q}
\end{align}
where \( N_2 \)  is positive definite, and~\( -N_3 \) is positive semidefinite.

The Hamiltonian density can be rewritten as a quadratic form associated with \( \hat{N} \):
\[
H = \frac{1}{2} \zeta \cdot \hat{I }\hat{N} \zeta.
\]
We find that \(\zeta\) has mixed dimensions: its first vector component corresponds to length, while the second corresponds to force divided by length. As a result, \(N_1\) is dimensionless, whereas \(N_3\) has the dimension of stress, and~\(N_2\) represents the inverse of stress (compliance).

Using the 6-by-6 constant matrix~\cite{ting1996anisotropic}
\begin{align}
    \hat{I}= \begin{bmatrix}
        0 & I \\
        I & 0 
    \end{bmatrix}
\end{align}
and in view of the symmetry of \(N_2\) and \(N_3\), one retrieved the fundamental symmetric matrix
\begin{align}
    \hat{I}\hat{N}= \begin{bmatrix}
        N_3 & N_1^T \\
        N_1 & N_2 
    \end{bmatrix}= (\hat{I}\hat{N})^T.
\end{align} According to~\cite{ting1996anisotropic}, \(N_2\) is positive definite, while \(-N_3\) is positive semidefinite. When seeking traveling wave solutions of the form \(\zeta = \Xi f(x_1 + p x_2)\), the problem reduces to a right eigenvalue problem.
\begin{align}
    \hat{N} \Xi = p \Xi
\end{align}
Substituting this into the total energy integral, we have:
\[
E = \int_\Sigma \frac{1}{2} \zeta \cdot  \hat{I} \hat{N} \zeta \, dx_1 dx_3.
\]

This representation highlights the symplectic nature of the system. The key property of the Hamiltonian is that it is independent of the ``time-like'' variable \(x_2\). This independence implies that the total energy \( E \) is conserved:
\[
\frac{dE}{dx_2} = 0.
\]

This conservation law can be derived from the canonical equations of motion and the symmetry of the fundamental matrix \(\hat{N} \). Specifically, the conservation of energy is expressed as:
\[
\int_\Sigma \zeta \cdot \hat{I} \hat{N}  \zeta \, dx_1 dx_3 = \text{constant},
\]
where \( \zeta = [\phi, u]^T \) is the vector of unknowns, and~\(\hat{N} \) is the fundamental symmetric matrix.

\subsection{Connection to the Stroh Formalism}
The Stroh formalism arises naturally in this framework. The evolution of the system is described by: 
\[
\frac{\partial}{\partial x_2} \zeta = \hat{N} \frac{\partial}{\partial x_1} \zeta,
\] 
where \( \hat{N} \) is the fundamental elasticity block matrix. \\ 

\section{Conclusions}\label{sec5}
 In this study, we integrated the incompressibility condition into Biot's poroelasticity framework to examine reversible (linear) poroelasticity. By adopting a Hamiltonian-based Stroh-like formulation, we aimed to improve the theoretical understanding of the system, particularly in conditions where no fluid pressure gradient exists.
\begin{itemize}
    \item The incorporation of the incompressibility condition into Biot’s poroelasticity allowed for the selection of appropriate energy–conjugate variable pairs, leading to a clearer description of the system.
    \item The Stroh-like formulation, grounded in Hamiltonian mechanics, provided a robust framework for analyzing poroelasticity under perfectly drained conditions, where dissipation is neglected.
 \item The absence of a fluid pressure gradient imposed constraints on the conjugate variables, reducing the degrees of freedom and offering a deeper insight into the reversible behavior of the system.
\item Our findings demonstrate the utility of the Hamiltonian formalism in theoretical studies and in designing experiments for systems where incompressibility constraints are relevant, ensuring an accurate assessment of mechanical properties and their coupling with fluid effects.
\end{itemize}
While this work focuses on the theoretical aspects of reversible poroelasticity under specific conditions, the application of these principles to more complex, real-world engineering problems remains an area for future exploration. Further research could investigate scenarios where dissipation and pressure gradients play a significant role, broadening the applicability of this framework.

\section*{Author Contributions}
Conceptualization, K.A. and V.T.; methodology, K.A. and V.T.; formal analysis, K.A. and V.T.; investigation, K.A.; writing---original draft preparation, K.A.; writing---review and editing, K.A. and V.T.; and supervision, V.T. All authors have read and agreed to the published version of the manuscript.

\section*{Funding}
{This} research was funded by NextGenerationEU project \textit{{{Transport} phonema in low-dimensional structures: models, simulations, and theoretical aspects}} (CUP D53D23006000006) and NextGenerationEU project \textit{Non-linear models for magma transport and volcanoes generation} (project code P20222B5P9).

\section*{Data Availability Statement}
No new data were created or analyzed in this study. Data sharing is not applicable to this article.

\section*{Acknowledgments}
The authors would like to thank the {{Gruppo} Nazionale per la Fisica Matematica} of the {Istituto Nazionale di Alta Matematica ``Francesco Severi''}.

\section*{Conflicts of Interest}
The authors declare no conflicts of interest.


\begin{thebibliography}{999}

\bibitem{stroh1962steady}
Stroh, A.N.
\newblock {Steady State Problems in Anisotropic Elasticity}.
\newblock {\em J. Math. Phys.} {\bf 1962}, {\em
41},~77--103. [\href{http://doi.org/10.1002/sapm196241177}{CrossRef}]

\bibitem{hwu2022stroh}
Hwu, C.; Becker, W.
\newblock Stroh formalism for various types of materials and deformations.
\newblock {\em J. Mech.} {\bf 2022}, {\em 38},~433--444. [\href{http://dx.doi.org/10.1093/jom/ufac031}{CrossRef}]

\bibitem{edmondson2009stroh}
Edmondson, R.; Fu, Y.
\newblock Stroh formulation for a generally constrained and pre-stressed
elastic material.
\newblock {\em Int. J. Non-Linear Mech.} {\bf 2009}, {\em
44},~530--537. [\href{http://dx.doi.org/10.1016/j.ijnonlinmec.2008.11.001}{CrossRef}]

\bibitem{chadwick1997application}
Chadwick, P.
\newblock The application of the Stroh formalism to prestressed elastic media.
\newblock {\em Math. Mech. Solids} {\bf 1997}, {\em
2},~379--403. [\href{http://dx.doi.org/10.1177/108128659700200402}{CrossRef}]

\bibitem{wang2024study}
Wang, X.; Zhou, J.; Chen, Q.; Huang, K.; Xue, Y.
\newblock Study on the attributes of transverse surface waves in a layered
piezoelectric semi-space with surface elasticity theory and Extended Stroh
formalism.
\newblock {\em Thin-Walled Struct.} {\bf 2024}, \emph{{202}
}, 112084. [\href{http://dx.doi.org/10.1016/j.tws.2024.112084}{CrossRef}]

\bibitem{qu2024defects}
Qu, Y.; Zhu, F.; Pan, E.
\newblock Defects and waves in anisotropic and layered media: Stroh formalism
in terms of the Cartesian system of vector functions.
\newblock {\em J. Mech.} {\bf 2024}, {\em 40},~687--697. [\href{http://dx.doi.org/10.1093/jom/ufae052}{CrossRef}]

\bibitem{wu2000extension}
Wu, K.C.
\newblock Extension of Stroh's formalism to self-similar problems in
two-dimensional elastodynamics.
\newblock {\em Proc. R. Soc. London. Ser. A Math. Phys. Eng. Sci.} {\bf 2000}, {\em
456},~869--890. [\href{http://dx.doi.org/10.1098/rspa.2000.0540}{CrossRef}]

\bibitem{ensam2023transmission}
Houyouk, J.; Manyo Manyo, J.A.; Ntamack, G.E.; Azrar, L. Transmission and Reflection and Coefficients of the Periodic Thermoelastic Multilayer Structures with Effect of Interface Using the Pseudo-Stroh Formalism. In \emph{Advances in Integrated Design and Production II, Proceedings of the 12th International Conference on Integrated Design and Production, CPI 2022, ENSAM, Rabat, Morocco, 10–12 May 2022}; Springer Nature: Cham, Switzerland, 2023; p. 273. [\href{http://dx.doi.org/10.1007/978-3-031-23615-0_28}{CrossRef}]

\bibitem{pan2023lamb}
Pan, E.; Zhu, F.; Li, N.; Qu, Y.
\newblock {Lamb waves in piezoelectric and flexoelectric plates: A new Stroh
formalism for gradient electro-mechanics.}
\newblock {2023}{.} 
\newblock [\href{http://dx.doi.org/10.2139/ssrn.4334620}{CrossRef}]

\bibitem{farm2016absorption}
F{\"a}rm, A.; Boij, S.; Glav, R.; Dazel, O.
\newblock Absorption of sound at a surface exposed to flow and temperature
gradients.
\newblock {\em Appl. Acoust.} {\bf 2016}, {\em 110},~33--42. [\href{http://dx.doi.org/10.1016/j.apacoust.2016.03.017}{CrossRef}]

\bibitem{wang2024characteristics}
Wang, L.; Yin, Z.Y.; Chen, W.
\newblock Characteristics of poroelastic deformation and dissipation of heat
and pore pressure in fluid-saturated porous strata.
\newblock {\em Comput. Geotech.} {\bf 2024}, {\em 167},~106046. [\href{http://dx.doi.org/10.1016/j.compgeo.2023.106046}{CrossRef}]

\bibitem{yang2017sound}
Yang, M.; Sheng, P.
\newblock Sound absorption structures: From porous media to acoustic
metamaterials.
\newblock {\em Annu. Rev. Mater. Res.} {\bf 2017}, {\em
47},~83--114. [\href{http://dx.doi.org/10.1146/annurev-matsci-070616-124032}{CrossRef}]

\bibitem{benjamin2024impacts}
Benjamin, G.; Neil, R.; Daniel, B.; Hamid, K.
\newblock Impacts of poroelastic spheres.
\newblock {\em arXiv} {\bf 2024}, arXiv:2411.05891. [\href{http://dx.doi.org/10.48550/arXiv.2411.05891}{CrossRef}]

\bibitem{corapcioglu1996propagation}
Corapcioglu, M.Y.; Tuncay, K.
\newblock Propagation of waves in porous media. In {\em Advances in Porous
Media}; Elsevier:  {Amsterdam, The Netherlands,} 1996; Volume~3, pp. 361--440. [\href{http://dx.doi.org/10.1016/S1873-975X(96)80007-2}{CrossRef}]

\bibitem{berryman2000elastic}
Berryman, J.G.; Wang, H.F.
\newblock Elastic wave propagation and attenuation in a double-porosity
dual-permeability medium.
\newblock {\em Int. J. Rock Mech. Min. Sci.}
{\bf 2000}, {\em 37},~63--78. [\href{http://dx.doi.org/10.1016/S1365-1609(99)00092-1}{CrossRef}]

\bibitem{pramanik2024theory}
Pramanik, D.; Manna, S.; Nobili, A.
\newblock Theory of elastic wave propagation in a fluid-saturated multi-porous
medium with multi-permeability.
\newblock {\em Proc. R. Soc. A} {\bf 2024}, {\em
480},~20230863. [\href{http://dx.doi.org/10.1098/rspa.2023.0863}{CrossRef}]

\bibitem{sharma2010piezoelectric}
Sharma, M.
\newblock Piezoelectric effect on the velocities of waves in an anisotropic
piezo-poroelastic medium.
\newblock {\em Proc. R. Soc. A Math. Phys. Eng. Sci.} {\bf 2010}, {\em 466},~1977--1992. [\href{http://dx.doi.org/10.1098/rspa.2009.0534}{CrossRef}]

\bibitem{chen2018biot}
Chen, H.; Gilbert, R.P.; Guyenne, P.
\newblock A biot model for the determination of material parameters of
cancellous bone from acoustic measurements.
\newblock {\em Inverse Probl.} {\bf 2018}, {\em 34},~085009. [\href{http://dx.doi.org/10.1088/1361-6420/aac520}{CrossRef}]

\bibitem{shao1998poroelastic}
Shao, J.
\newblock Poroelastic behaviour of brittle rock materials with anisotropic
damage.
\newblock {\em Mech. Mater.} {\bf 1998}, {\em 30},~41--53. [\href{http://dx.doi.org/10.1016/S0167-6636(98)00025-8}{CrossRef}]

\bibitem{yang2024modified}
Yang, C.; Wang, Y.; Stovas, A.; Wang, L.
\newblock Modified Biot elastic coefficients in poroelastic solid.
\newblock {\em J. Earth Sci.} {\bf 2024}, {\em 35},~1397--1401. [\href{http://dx.doi.org/10.1007/s12583-024-1994-8}{CrossRef}]

\bibitem{abdollahipour2016complete}
Abdollahipour, A.; Marji, M.F.; Bafghi, A.Y.; Gholamnejad, J.
\newblock A complete formulation of an indirect boundary element method for
poroelastic rocks.
\newblock {\em Comput. Geotech.} {\bf 2016}, {\em 74},~15--25. [\href{http://dx.doi.org/10.1016/j.compgeo.2015.12.011}{CrossRef}]

\bibitem{barnett2000bulk}
Barnett, D.M.
\newblock Bulk, surface, and interfacial waves in anisotropic linear elastic
solids.
\newblock {\em Int. J. Solids Struct.} {\bf 2000}, {\em
37},~45--54. [\href{http://dx.doi.org/10.1016/S0020-7683(99)00076-1}{CrossRef}]

\bibitem{fu2007hamiltonian}
Fu, Y.
\newblock Hamiltonian interpretation of the {S}troh formalism in anisotropic
elasticity.
\newblock {\em Proc. R. Soc. A Math. Phys. Eng. Sci.} {\bf 2007}, {\em 463},~3073--3087. [\href{http://dx.doi.org/10.1098/rspa.2007.0093}{CrossRef}]

\bibitem{nobili2022hamiltonian}
Nobili, A.; Radi, E.
\newblock Hamiltonian/Stroh formalism for anisotropic media with
microstructure.
\newblock {\em Philos. Trans. R. Soc. A} {\bf 2022},
{\em 380},~20210374. [\href{http://dx.doi.org/10.1098/rsta.2021.0374}{CrossRef}] [\href{http://www.ncbi.nlm.nih.gov/pubmed/35858079}{PubMed}]

\bibitem{nobili2024hamiltonian}
Nobili, A.
\newblock Hamiltonian/{S}troh formalism for reversible poroelasticity (and
thermoelasticity).
\newblock {\em Int. J. Solids Struct.} {\bf 2024}, {\em
301}, {112935}. [\href{http://dx.doi.org/10.1016/j.ijsolstr.2024.112935}{CrossRef}]

\bibitem{nobili2021quasi}
Nobili, A.; Pichugin, A.
\newblock Quasi-adiabatic approximation for thermoelastic surface waves in
orthorhombic solids.
\newblock {\em Int. J. Eng. Sci.} {\bf 2021}, {\em
161},~103464. [\href{http://dx.doi.org/10.1016/j.ijengsci.2021.103464}{CrossRef}]

\bibitem{biot1962mechanics}
Biot, M.
\newblock Mechanics of deformation and acoustic propagation in porous media.
\newblock {\em J. Appl. Phys.} {\bf 1962}, {\em 33},~1482--1498. [\href{http://dx.doi.org/10.1063/1.1728759}{CrossRef}]

\bibitem{berger1999stabilized}
Berger, L.; Bordas, R.; Kay, D.; Tavener, S.
\newblock A stabilized finite element method for nonlinear poroelasticity.
\newblock {\em Comput. Methods Appl. Mech. Eng.} {\bf
1999}, {\em 174},~299--317.

\bibitem{abousleiman2002porothermoelasticity}
Abousleiman, Y.; Ekbote, S.
\newblock Porothermoelasticity in transversely isotropic porous materials.
\newblock In Proceedings of the IUTAM Symposium on Theoretical and Numerical
Methods in Continuum Mechanics of Porous Materials, {Stuttgart, Germany, 5--10 September 1999}; Springer: {Berlin/Heidelberg, Germany,} 2002; pp. 145--152. [\href{http://dx.doi.org/10.1007/0-306-46953-7}{CrossRef}]

\bibitem{schanz2009poroelastodynamics}
Schanz, M.
\newblock Poroelastodynamics: {L}inear models, analytical solutions, and
numerical methods.
\newblock {\em Appl. Mech. Rev.} {\bf 2009}, {\em 62},~1--15. [\href{http://dx.doi.org/10.1115/1.3090831}{CrossRef}]

\bibitem{biot1957elastic}
Biot, M.A.; Willis, D.G.
\newblock The elastic coefficients of the theory of consolidation. \emph{J. Appl. Mech.} {\bf 1957}, {\emph{24}}, {594--601.}
\newblock [\href{http://dx.doi.org/10.1115/1.4011606}{CrossRef}]

\bibitem{biot1955theory}
Biot, M.
\newblock Theory of elasticity and consolidation for a porous anisotropic
solid.
\newblock {\em J. Appl. Phys.} {\bf 1955}, {\em 26},~182--185. [\href{http://dx.doi.org/10.1063/1.1721956}{CrossRef}]

\bibitem{ting1996anisotropic}
Ting, T.C.
\newblock {\em Anisotropic Elasticity: Theory and Applications}; Oxford University Press: {Oxford, UK,} 1996; Volume~45.

\end{thebibliography}
\end{document}